# Single-step Estimation of Absolute Technical Efficiency Indices

Montacer BEN CHEIKH LARBI, Ph.D.

La Rochelle Université, France

ORCID id: 0009-0002-0349-5224

mbenchei@univ-lr.fr

Sina BELKHIRIA, Ph.D.

Institut Supérieur de Gestion de Tunis, Université de Tunis, Tunisia

ORCID id : 0009-0009-4136-8374

sina.belkhiria@isg.u-tunis.tn

**Abstract**

*Technical efficiency indices (TEIs) can be estimated using the traditional stochastic frontier analysis approach, which yields relative indices that do not allow self-interpretations. In this paper, we introduce a single-step estimation procedure for TEIs that eliminates the need to identify best practices and avoids imposing restrictive hypotheses on the error term. The resulting indices are absolute and allow for individual interpretation. In our model, we estimate a distance function using the inverse coefficient of resource utilization, rather than treating it as unobservable. We employ a Tobit model with a translog distance function as our econometric framework. Applying this model to a sample of 19 airline companies from 2012 to 2021, we find that:*

*(1) Absolute technical efficiency varies considerably between companies with medium-haul European airlines being technically the most efficient, while Asian airlines are the least efficient;*

*(2) Our estimated TEIs are consistent with the observed data with a decline in efficiency especially during the Covid-19 crisis and Brexit period;*

*(3) All airlines contained in our sample would be able to increase their average technical efficiency by 0.209% if they reduced their average kerosene consumption by 1%;*

*(4) Total factor productivity (TFP) growth slowed between 2013 and 2019 due to a decrease in Disembodied Technical Change (DTC) and a small effect from Scale Economies (SE). Toward the end of our study period, TFP growth seemed increasingly driven by the SE effect, with a sharp decline in 2020 followed by an equally sharp recovery in 2021 for most airlines.*

# 1. Introduction

All variants of the stochastic frontier analysis method require two-step techniques for estimating Farrell-Debreu technical efficiency indices (TEIs). In the first step, a model is estimated by maximum likelihood. In the second step, maximum likelihood residuals are decomposed into a nonnegative error term from which TEIs are derived, and a noise component. TEIs obtained in this manner are sensitive to the distribution assigned to the nonnegative error term. The values of these indices are not relevant; only their ranking is. Therefore, these indices are relative and do not have specific or individual interpretations.

This research aims to introduce a new parametric approach for estimating TEIs. In this paper, we propose a one-step estimation procedure for TEIs which relies on estimating an input-oriented distance function whose endogenous variable is measured by the inverse of the coefficient of resource utilization.

The distance function has been widely used since the early 2000s to assess technical efficiency and was originally conceived because (i) it delivers a primal representation of the production technology, i.e., it does not require any knowledge of input prices, and (ii) unlike the familiar production function it is capable of representing multi-output technologies.

Empirical studies have followed the recommendations by Färe and Grosskopf (1990) that the distance should be set to one because it is unobservable. Then, an input or output is arbitrarily chosen as the endogenous model variable depending on whether the distance function is input- or output-oriented. However, this approach ignores the work of Debreu (1951), who stated that the distance should be the inverse of the coefficient of resource utilization. Thus, if we have a measure of the rate of capacity utilization, we can estimate the distance function without having to impose any hypothesis on the value of the endogenous variable.

In this paper, we use the load factor as a proxy for capacity utilization. In the past, the load factor has been used only sporadically as an argument of the distance function, mainly in the Malmquist index models. For example, Huang et al. (2020) evaluated the productivity change for 15 airline companies using a Malmquist productivity index model, which is given by the ratio of distance functions. Considering that the load factor is neither an input nor an output, they introduced it as an attribute of the distance function.

Since the load factor is truncated and must be positive without exceeding 1, our econometric framework corresponds to a Tobit regression model. The efficiency indices are obtained from the inverse of the estimated distance function. Unlike indices obtained by the two-step techniques, our indices are absolute. Their values indicate the level of technical performance of companies. The proposed technique does not require searching for a benchmark, nor does it necessitate an asymmetric error term. This means that it does not require an individual dimension and can be applied on time series data.

In the SFA literature, individual effects are used to predict efficiency, and the main differences between the existing models are largely due to differences in the assumptions about the distribution of the individual effects. When panel data are available, individual effects are compelled to be time-dependent in order to make the efficiency indices time-varying. The models proposed by Cornwell et al. (1990), Kumbhakar (1990), Battese and Coelli (1992), and Lee and Schmidt (1993) are commonly used in empirical studies to estimate time-varying technical efficiency. In this study, we do not link TEIs to individual effects. The inverse of our estimated distance function provides time-varying TEIs in accordance with production economics theory.

Tobit models have been widely used when efficiency indices are mainly derived by a Data Envelopment Analysis (DEA) approach. Firstly, the so-called two-stage DEA approach calculates technical efficiency scores using a linear program. Then, these scores are regressed on a set of explanatory variables to assess the determinants of technical efficiency. When regressing the TEIs on a set of variables, we must remember that the arguments of TEIs are the outputs, $y$, and the inputs, $x$, and we must ensure that the technical efficiency function is homogeneous of degree $-1$ in $x$. Surprisingly, to the best of our knowledge, these regularity conditions of the $\text{TEI}(y, x)$ function have never been imposed in the literature.

Merkert and Hensher (2011) evaluated the determinants of efficiency scores in a two-stage DEA framework. None of the inputs and outputs they used in their DEA model appear in their second-stage Tobit regression. Xu et al. (2021) used both desirable and undesirable outputs in their DEA model, but again the outputs and inputs are not part of their Tobit regression. Ngo and Tsui (2022) used a two-stage DEA procedure to derive TEIs for Asia-Pacific airlines using three inputs and three outputs. In their Tobit model, there is no output or input on the right-hand side of the equation. In addition, they use fuel prices as an explanatory variable of TEIs. This is not consistent with the theory of production since the technical efficiency function is a primal representation of the production technology. The empirical results of the three studies mentioned above show that the homogeneity condition of the efficiency function has not been imposed.

The method we propose here to estimate the distance function has the advantage that it is easy to decompose total factor productivity and interpret the results. Tsionas et al. (2017) and Huang et al. (2020) contain excellent literature reviews on technical efficiency evaluation and productivity for air carriers. They show that most efficiency and productivity studies are conducted using the DEA approach. Empirical studies that employ the SFA approach are based on production or cost functions; if they use distance functions, then distances must be unitary. This is the case in the model of Tsionas et al. (2017), where their output-oriented distance function is, in fact, a transformation function (Fuss and McFadden, 1978).

The remainder of this paper is organized as follows. Section 2 examines the methodology. The results are discussed in section 3, while section 4 presents the conclusions of the study and provides an outlook on possible future research directions.

## 2. Methodology and model Specification

### 2.1. The Input Distance Function and Absolute Technical Efficiency Indices

Aigner et al. (1977) and Meeusen and van den Broeck (1977) introduced stochastic production frontier models in which the error term consists of two components: (1) the usual disturbance and (2) a measure of technical inefficiency in the production process. This model is given by the following relation: $y = f(x; \beta) + v - u$, where $y$ is the output, $f$ the functional form of the production function, $x$ a vector of inputs, $\beta$ unknown parameters, $v$ the two-sided usual disturbance, and $u$ a nonnegative technical inefficiency term.

This specification has been widely used in the literature and has had important empirical applications since the early 1980s. It has also been the subject of various developments, notably concerning the assumptions made for the distribution of the inefficiency term $u$.

When dealing with panel data, the extensions were mainly related to time-varying efficiency indices. In this context, the model is written as $y_{it} = f(x_{it}; \beta)\exp(u_{it})$, which immediately shows why the functional forms used in the SFA approach must be logarithmic (Cobb-Douglas or Translog).

When the distance function first began to be explored in empirical work as a primal representation of production technology, it was thought that distance, *D*, was not observable, and the recommended solution was to set $D = 1$ (Färe and Grosskopf, 1990; Kumbhakar and Knox Lovell, 2000; Coelli and Pelerman, 2000), meaning that firms are technically efficient. Next, an input is arbitrarily chosen and placed in the left-hand side of the equation. The explanatory variables are then correlated with the error term, which violates the basic assumptions of the stochastic frontier model and leads to biased estimators. Finally, deviations from the efficiency frontier are obtained through the nonnegative error term, $u$, in the same way as in production frontier models.

Here, we do not adopt this technique but propose a single-step procedure to derive TEIs by estimating an input-oriented distance function. We define distance as the inverse of the capacity utilization rate in accordance with Debreu's (1951) definition of the coefficient of resource utilization.

The input-oriented distance function is defined as:

$$D(y, x) = \max\{\lambda > 0, x/\lambda \in L(y)\}$$

where $y$ is a vector of outputs, $x$ is a vector of inputs, and $L(y)$ is the input requirement set. Following Diewert (1982), the input-oriented distance function is interpreted as the biggest number that will deflate the input, $x$, onto the boundary of the input requirement set $L(y)$.

In Fig. 1, the input-oriented distance function is given by $D(y, x) = \parallel x \parallel / \parallel x^* \parallel$, where $\parallel x \parallel = \sqrt{x_1^2 + x_2^2 + \cdots + x_k^2}$ is the Euclidian distance of $x$. Hence, if $x$ is technically efficient, i.e., if $x$ is on the isoquant $Q(y)$, then $D(y, x) = 1$. We obtain $D(y, x) = [\text{TEI}(y, x)]^{-1}$, where $\text{TEI}(y, x)$ is the input-oriented measure of technical efficiency.

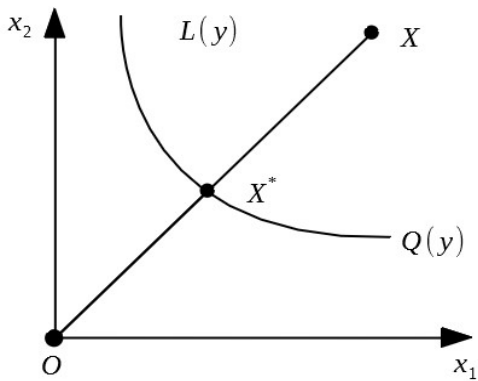

Figure 1. Illustrating the input distance function.

It is important to note that since we measure technical efficiency as inverse distance, we do not need to impose any restrictions on the choice of the functional form of the distance function.

In this paper, we suppose that the production technology is represented by a translog functional form. Our model can be written as follows:

$$\ln(D_{it}) = \ln(LF_{it}^{-1}) = \beta_0 + \sum_j \beta_j \ln x_{jit} + \beta_y \ln y_{it} + \beta_t t_{it} + .5 \sum_j \sum_k \beta_{jk} \ln x_{jit} \ln x_{kit} +$$
$$\sum_j \beta_{jy} \ln x_j \ln y_{it} + \sum_j \beta_{jt} \ln x_j t_{it} + .5 \beta_{yy} (\ln y_{it})^2 + \beta_{ty} t \ln y_{it} + .5 \beta_{tt} t_{it}^2 + \sum_i \gamma_i z_i + \delta_B z_B +$$
$$+ \delta_C z_C + v_{it}, \qquad (1)$$

where the indices $i$ and $t$ denote the individual and time, respectively, $LF_{it}$ is the load factor, $x_{jit}$ the $j$th input, $y_{it}$ the output, $t_{it}$ is a time trend that represents, $z_i$ are individual dummies, $z_B$ is a dummy variable representing the Brexit effect, taking the value of one for the years 2017-2019 if the carrier is European, and zero otherwise, $z_C$ is a dummy variable for the Covid effect, taking the value of one for 2020 and 2021, and zero otherwise, and $v_{it}$ an error term.

In order for Eq. (1) to be well-behaved, we impose the following regularity conditions:

$$\sum_j \beta_j = 1, \quad \sum_j \beta_{jk} = \sum_j \beta_{jy} = \sum_j \beta_{jt} = 0, \quad \beta_{jk} = \beta_{kj}, \quad i \neq j. \qquad (2)$$

Since heteroscedasticity can have a significant effect on the estimated TEIs, we suppose that $v_{it} \sim iid. N(0, \sigma_{it}^2)$ and correct for heteroscedasticity by assuming the following multiplicative pattern for the variance of the error term: $\sigma_{it}^2 = \exp(X_{it} \alpha)$, where $X_{it}$ are the arguments of the distance function and $\alpha$ is a vector of unknown parameters.

Based on the above model, we can calculate the rate of technical change as:

$$TC = \partial \ln D / \partial t = - \partial \ln TE / \partial t = \beta_t + \beta_{tt} t + \sum_j \beta_{jt} \ln x_j + \beta_{ty} \ln y \qquad (3)$$

Finally, following Baltagi and Griffin (1988), we can estimate total factor productivity growth, $TFPG$, by totally differentiating Equation (1), which yields:

$$TFPG_{it} = TC_{it} + (1 - \varepsilon_{Dy}) \ln(y_{it} / y_{i,t-1}) \qquad (4)$$

where $\varepsilon_{Dy} = -\partial \ln D / \partial \ln y$.

Equations (3) and (4) show that $TFPG$ has four components: Disembodied technical change: $DTC = \beta_t + \beta_{tt} t$, embodied technical change: $ETC = \sum_j \beta_{jt} \ln x_j$, scale technical change: $STC = \beta_{ty} \ln y$, and scale economy: $SE = (1 - \varepsilon_{Dy}) \ln(y_{it} / y_{i,t-1})$.

If returns to scale are constant, i.e., if $RTS = \varepsilon_{Dy}^{-1} = 1$, then the $TFPG$ is reduced to its technical change component.

## 2.2. Data

We analyzed 19 of the largest passenger airlines (listed in figure 1 and Table 4) over the period 2012 to 2021, building our panel database from the airlines' published annual reports. The variables considered in the subsequent analysis include:

- Passenger Load Factor        $(LF)$,
- Total Operating Revenue      $(y)$,
- Depreciation and Amortization $(K)$,
- Number of Employees          $(L)$,
- Jet Fuel Consumption         $(E)$.

As mentioned above, we use the inverse load factor as a proxy for distance, $LF^{-1} = D$. The output, $y$, is measured by the total operating revenue. The three inputs we use in this study are capital $(K)$, labor $(L)$, and energy $(E)$, which are measured by depreciation and amortization, the number of employees, and jet fuel consumption, respectively.

When airlines measured jet fuel consumption in kilotons or thousands of liters, we converted it to millions of gallons.

Table 1 depicts the descriptive statistics for our variables and their respective units.

**Table 1.** Descriptive Statistics

| Variables | N | Mean | Std. Dev. | Min | Max | Units/Definition |
|---|---|---|---|---|---|---|
| $LF$ | 190 | 0.78 | 0.11 | 0.14 | 0.96 | The available seating capacity filled with passengers |
| $Y$ | 190 | 18,531.09 | 11,517.85 | 1,515.24 | 43,579.64 | Million US$ |
| $K$ | 190 | 1,584.65 | 910.61 | 161.65 | 5,224.79 | Million US$ |
| $L$ | 190 | 57,850.89 | 34,943.76 | 7,840 | 137,784 | Full-time equivalent |
| E | 190 | 1,980.30 | 1,139.58 | 166.01 | 4,537 | Million gallons |

*Note: Table 1 exhibits the definition and the units of variables and the descriptive statistics for the database which is composed of 19 airline companies from 2012 to 2021. **Std.Dev.**: Standard deviation, **Min**: Minimum, **Max**: Maximum.*

## 3. Results and Discussion

### 3.1. Model Specification Tests

The results from three specification tests for the distance function are summarized in Table 2.

The first hypothesis tests whether the translog functional form of the distance function could be reduced to a Cobb-Douglas functional form, using the following null hypothesis: $H_{01}: \beta_{jk} = \beta_{jy} = \beta_{yy} = \beta_{ty} = \beta_{tt} = 0, \forall j, k$. The second hypothesis tests whether the returns to scale are constant with the null hypothesis given by: $H_{02}: \beta_y = -1, \beta_{jy} = \beta_{yy} = \beta_{ty} = 0, \forall j$. The third hypothesis tests the existence of technical change with the null hypothesis given by: $H_{03}: \beta_t = \beta_{jt} = \beta_{ty} = \beta_{tt} = \delta_B = \delta_C = 0, \forall j$.

**Table 2.** Model specification tests.

| Null hypothesis | Statistic | Pr>Chi-square | Decision |
|---|---|---|---|
| Test 1: Cobb-Douglas Functional Form | 88.24 | <0.0001 | Reject |
| Test 2: Constant Returns to Scale | 252.73 | <0.0001 | Reject |
| Test 3: No Technical Change | 425.82 | <0.0001 | Reject |

***Note:*** *Table 2 provides the results of three specification tests for the distance function. Test 1 concerns the functional form (Cobb-Douglas vs translog), test 2 is about the constancy of returns to scale and test 3 deals with the existence of technical change.*

### 3.2. Estimating Absolute Efficiency Indices

The empirical results of the translog distance function, estimated by maximum likelihood, are presented in Table 3.

**Table 3.** Maximum likelihood estimates of the model.

| Parameter | Coefficient | P-Value | Parameter | Coefficient | P-Value |
|---|---|---|---|---|---|
| Constant | 0.297336 | <.0001 | z1 | 0.561021 | <.0001 |
| $\ln y$ | -0.593745 | <.0001 | z2 | 0.565005 | <.0001 |
| $\ln K$ | 0.204451 | <.0001 | z3 | -0.184319 | 0.0005 |
| $\ln L$ | 0.685113 | <.0001 | z4 | -0.113492 | 0.0017 |
| $\ln E$ | 0.110436 | 0.0595 | z5 | 0.683347 | <.0001 |
| $.5 (\ln y)^2$ | 0.052062 | 0.2905 | z6 | -0.173345 | <.0001 |
| $\ln K \ln y$ | 0.039240 | 0.2775 | z7 | -0.172615 | 0.0035 |
| $\ln L \ln y$ | 0.217154 | <.0001 | z8 | -0.251890 | <.0001 |
| $\ln E \ln y$ | -0.256393 | <.0001 | z9 | 0.066778 | <.0001 |
| $.5 (\ln K)^2$ | 0.231760 | 0.0023 | z10 | 0.649484 | <.0001 |
| $\ln K \ln L$ | 0.247746 | 0.0002 | z11 | 0.152897 | <.0001 |
| $\ln K \ln E$ | -0.479506 | <.0001 | z12 | 0.070368 | 0.2224 |
| $.5 (\ln L)^2$ | -0.242152 | 0.0253 | z13 | -0.180902 | 0.0004 |
| $\ln L \ln E$ | -0.005594 | 0.9472 | z14 | 0.402020 | <.0001 |
| $.5 (\ln E)^2$ | 0.485100 | <.0001 | z15 | 0.492576 | <.0001 |
| $t$ | -0.032827 | <.0001 | z16 | 0.400680 | <.0001 |
| $t \ln y$ | 0.007825 | 0.0044 | z17 | 0.197729 | <.0001 |
| $t \ln K$ | -0.002760 | 0.5647 | z18 | 0.300541 | <.0001 |
| $t \ln L$ | -0.012648 | 0.0243 | Brexit | -0.042064 | 0.0218 |
| $t \ln E$ | 0.015408 | 0.0043 | Covid | 0.134461 | 0.0005 |

***Note:*** *Table 3 reports the empirical results of the translog distance function estimated by maximum likelihood.*

To avoid the singularity of the hessian matrix and convergence problems, the output and the inputs were divided by their respective averages and the squared trend was excluded from the model. Most of the other parameters are highly significant. The two dummy variables for covid and Brexit are also significant.

The estimated TEIs obtained from the inverse of the estimated distance function are shown in Fig. 2 and summarized in Table 4 by their individual average values.

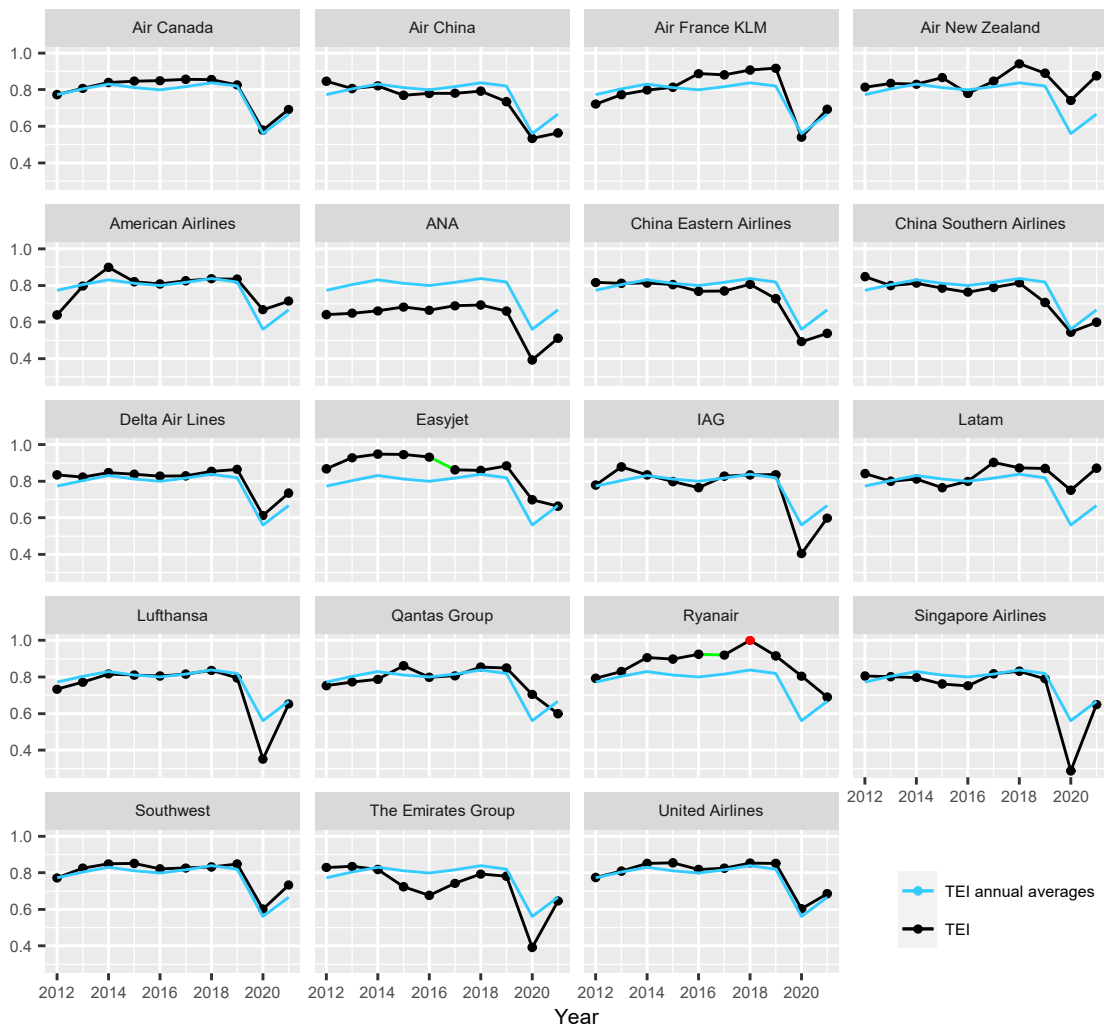

**Fig. 2.** *Estimated technical efficiency (unitary indices are highlighted in red; the 2016-2017 decline, only seen in two European airlines and likely due to Brexit, is marked in green)*

**Table 4.** Average Technical efficiency per company (2012-2021)

| Company | Average TEI |
| --- | --- |
| Ryanair | 0.86809 |
| Easyjet | 0.85890 |
| Air New Zealand | 0.84178 |
| Latam | 0.82838 |
| Delta Air Lines | 0.80663 |

| Company | Average TEI |
|---|---|
| Southwest | 0.79596 |
| Air France KLM | 0.79293 |
| United Airlines | 0.79260 |
| Air Canada | 0.79233 |
| American Airlines | 0.78409 |
| Qantas Group | 0.77871 |
| IAG | 0.75563 |
| China Southern Airlines | 0.74618 |
| Air China | 0.74260 |
| Lufthansa | 0.73873 |
| China Eastern Airlines | 0.73497 |
| Singapore Airlines | 0.72948 |
| The Emirates Group | 0.72362 |
| ANA | 0.62408 |

***Note:*** *Table 4 reports the Average estimated Technical Efficiency indices* obtained from the inverse of the estimated distance function

By examining estimated TEIs, we can make four observations. Firstly, the TEIs exhibit non-linear patterns. This is a consequence of the estimation procedure proposed here, which does not require the search for a benchmark at each period nor does it impose restrictive assumptions on the error term. As a result, these indices are absolute and not relative (measured with respect to a benchmark) like the indices commonly proposed in the literature. Thus, each index can be interpreted individually with regard to its proximity to the isoquant. Our indices thus reflect the nature of the data and are not constrained by any model assumptions. This aspect is particularly perceptible for the year 2020, the peak of the Covid-19 crisis, where Fig. 2 shows a sharp decline in the TEIs for all airlines.

Secondly, throughout the sample period, the top two efficient companies consistently exceeded the individual averages. This result supports the paradigm that technical efficiency should not be viewed as an isolated performance indicator but rather as the result of an accumulation of managerial and organizational know-how and experience acquired over time that makes the firm more competitive. The least efficient company consistently fell short of the individual averages. The Emirates Group, Singapore Airlines, and the three Chinese companies show a downward trend in their indices and are among the least efficient companies.

Thirdly, the Covid-19 crisis had a significant negative impact on the technical efficiency of all companies. If we were to estimate the TEIs by any approach available in the literature, we would not be able to observe this decline in technical efficiency for the 2020 benchmarks. Nevertheless, the Covid-19 pandemic led to huge drops in revenue for all airlines, partly due to the successive lockdowns and partly because of reduced demand following the imposed health restrictions. According to the International Air Transport Association (IATA) Economics Report (2020), the growth rate of air transport, as measured by either revenue passenger kilometers (RPK) or the revenue passenger kilometer index, declined by more than 65.9% in 2020. This loss had an absolute impact on the efficiency of airlines included in our sample which is reflected as a decline in indices in 2020 for all companies (Fig. 2).

Fourthly, Ryanair and EasyJet primarily operate in Europe. So, the decline in their TEIs in 2017 is most likely attributable to Brexit. The revenue-weighted average of these companies' TEIs declined by 3.95% between 2016 and 2017. These declines (highlighted in green in Fig.2) are slightly offset by those generated by the Covid-19 crisis between 2019 and 2020.

In addition, we have estimated the elasticities of the TEIs with respect to inputs and output in order to assess the impact of their utilization on the performance of the companies. Estimating these elasticities is straightforward and can be achieved by:

$$\partial \ln(\text{TEI})/\partial \ln x_i = -\partial \ln D/\partial \ln x_i \qquad (5)$$

$$\partial \ln(\text{TEI})/\partial \ln y = -\partial \ln D/\partial \ln y = 1/RTS \qquad (6)$$

The time-averaged elasticities of the TEIs with respect to capital and labor are -0.15 and -0.64, respectively. While these averages have the expected signs, results for individual companies are rather variable over time (Fig. 3). For energy, the average elasticity is -0.21, indicating that the airlines included in our sample could increase their average efficiency by 0.209% if they reduced their fuel consumption by 1%. This could be accomplished by decreasing the average age of the fleet or by acquiring more fuel-efficient aircrafts. Finally, every company has the potential to enhance its TEIs by boosting output, albeit the magnitude of this improvement remains relatively modest given our findings that they operate under increasing returns to scale.

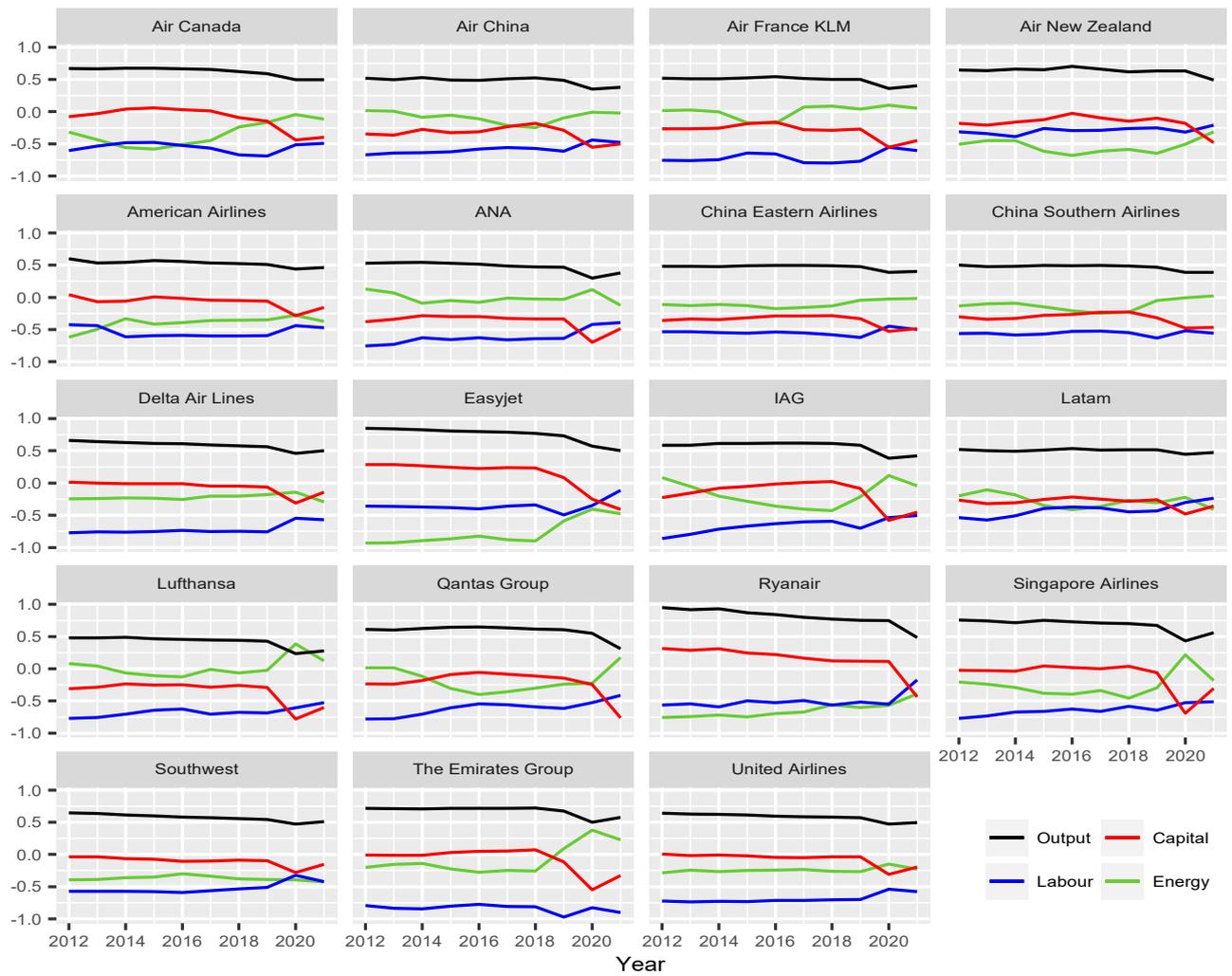

***Fig 3.*** *Elasticities of Technical Efficiency Indices with respect to output and inputs.*

### 3.3. Decomposition of Total Factor Productivity Growth

By examining the estimations of *TFP* growth rates and their components (Fig. 4), we note that technical change exhibits a negative trend across all companies throughout the entire period. This is attributed to the consistent nature of *DTC* in our model ($\beta_t = -0.033$), which holds a higher absolute value compared to the combined values of *SCT* and *ETC*. Moreover, *SE* demonstrates significant variability among companies, notably during the 2013-2019 period, preceding the onset of the COVID crisis. Furthermore, *TFPG* shows a strong correlation with *SE* for each company, with correlation coefficients exceeding 0.99.

In 2020, most companies reduced their workforce, some by more than 20%, but this reduction couldn't offset the output decline. That's why the TFP slowdown was particularly strong in 2020.

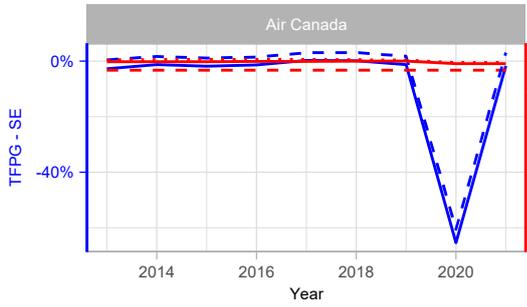
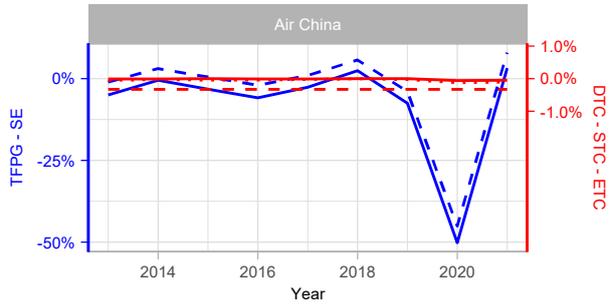
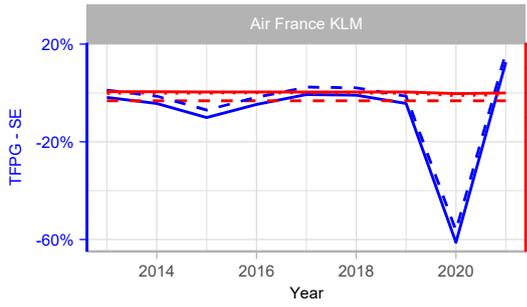
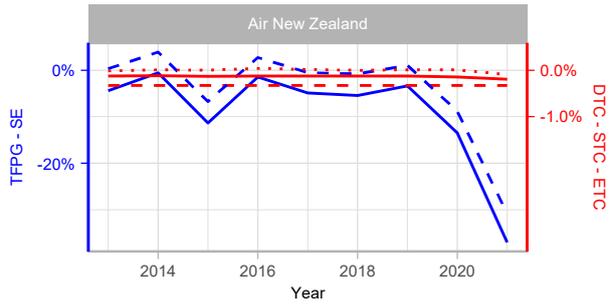
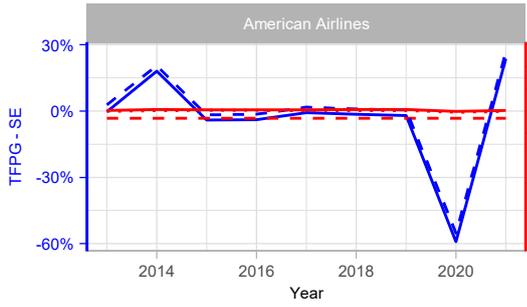
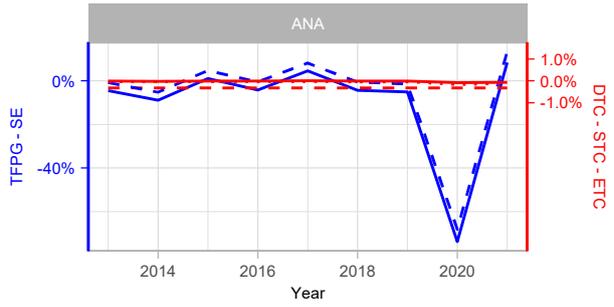
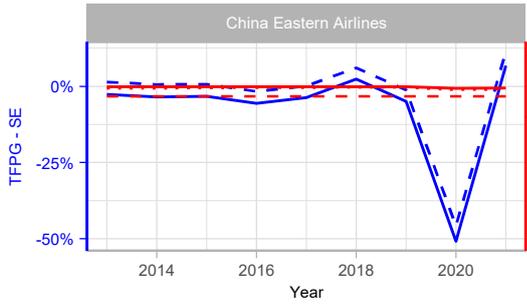
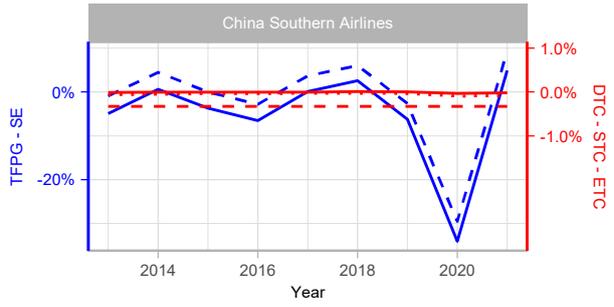
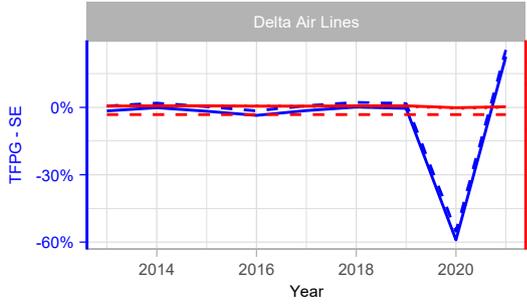
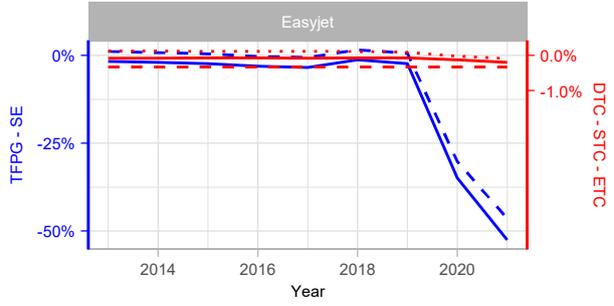

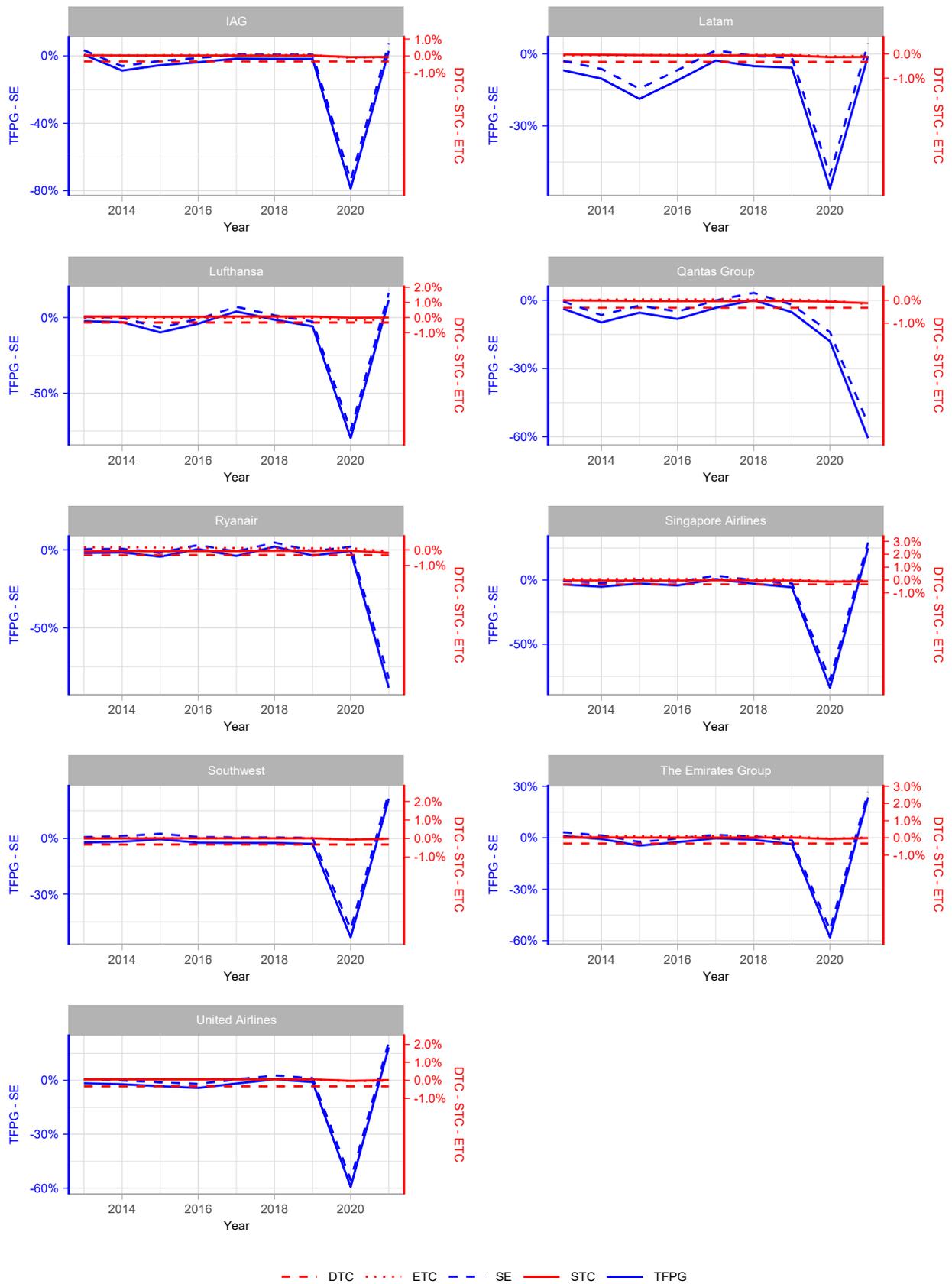

**Fig. 4.** *The decomposition of total factor productivity growth (TFPG).*

## 4. Conclusion

In this study, we introduced a new parametric approach for estimating TEIs. Our procedure has several theoretical and practical advantages over traditional approaches available in the literature. From a theoretical point of view, our procedure is consistent with production economic theory since it measures the distance from the inverse of the coefficient of resource utilization, which does not need to be unitary. As a result, our TEIs, including those of the benchmarks, are amenable to individual and specific interpretations.

In practical terms, our procedure enables the estimation of TEIs in a single step without the need to identify benchmarks or impose restrictive hypotheses on the error term. By eliminating the best practice identification step, our method becomes applicable to time series analysis. Furthermore, without basing the estimation of TEIs on individual effects, we can avoid imposing strong assumptions on the error term.

Finally, the functional form can be freely chosen in the sense that it is no longer necessary to use Cobb-Douglas or translog-type models.

Our TEIs reflect the nature of the data and are not constrained by any model assumptions. This aspect is particularly perceptible for the year 2020 when indices show a sharp decline due to Covid-19 crisis and a slight fall in 2017 which is most likely attributable to Brexit for European carriers, which demonstrates that their values exclusively depend on a company's productive performance.

Future research could focus on applying the procedure proposed here to functional forms other than translog and Cobb-Douglas. Furthermore, we believe that it would be useful to test the validity of the static equilibrium hypothesis and possibly estimate a temporary equilibrium model using a short-run distance function given the presence of economic crises over our study period, which could make capital quasi-fixed. The main limitation of our study concerns the absence of a non-frontier approach to the measurement of TFP growth. A translog Divisia index should be calculated and compared to the TFP growth obtained through the parametric approach. Alternatively, future studies could focus on a direct comparison of TEIs obtained by more common methods and those obtained through our approach.

*Acknowledgements:* None.
*Funding: This research did not receive any specific grant from funding agencies in the public, commercial, or not-for-profit sectors.*
*Declarations of interest:* None.

# References

Aigner, D., Knox Lovell, C.A., Schmidt, P., 1977. Formulation and estimation of stochastic frontier production function models. Journal of Econometrics, Volume 6, Issue 1, July 1977, Pages 21-37. https://doi.org/10.1016/0304-4076(77)90052-5.

Baltagi, B.H., Griffin, J.M., 1988. A General Index of Technical Change. Journal of Political Economy 96, 20-41. http://dx.doi.org/10.2307/2526831.

Battese, G.E., Coelli, T.J., 1992. Frontier production functions, technical efficiency, and panel data: With application to paddy farmers in India. Journal of Productivity Analysis 3(1):153-169. June 1992. DOI:10.1007/BF00158774.

Coelli, T., Pelerman, S., 2000. Technical Efficiency of European Railways: A Distance Function Approach. Applied Economics, 32(15):1967-76. DOI:10.1080/00036840050155896.

Cornwell, C.M., Schmidt, P., Sickles, R.C., 1990. Production frontiers with cross-sectional and time series variation in efficiency levels. Journal of Econometrics, 46(1):185-200. DOI:10.1016/0304-4076(90)90054-W.

Debreu, G., 1951. The coefficient of resource utilization. Econometrica, 19(3), July, 273-292. http://dx.doi.org/10.2307/1906814.

Diewert, W.E., 1982. Duality approaches to microeconomic theory. Handbook of mathematical economics 2, Chapter 12. 535-599.

Färe R., Grosskopf, S., 1990. A distance function approach to price efficiency. Journal of Public Economics, 43(1):123-126. North-Holland (1990). https://doi.org/10.1016/0047-2727(90)90054-L.

Fuss M., McFadden D., 1978. Production economics: A dual approach to theory and applications. Amsterdam: North Holland, 1978.

Huang F., Zhou D., Hu J.-L., Wang Q., 2020. Integrated airline productivity performance evaluation with $CO_2$ emissions and flight delays. Journal of Air Transport Management. 84(6):101770. https://doi.org/10.1016/j.jairtraman.2020.101770.

International Air Transport Association (IATA) Economics Report, 2020. Air Passenger Market Analysis. December 2020. IATA, Geneva. https://www.iata.org/en/iata-repository/publications/economic-reports/air-passenger-monthly-analysis---december-2020/.

Kumbhakar, S.C., 1990. Production frontiers, panel data, and time-varying technical inefficiency. Journal of Econometrics, 46(1-2):201-212. http://dx.doi.org/10.1016/0304-4076(90)90055-x.

Kumbhakar, S.C., Knox Lovell, C.A., 2000. Stochastic Frontier Analysis. Cambridge University Press. https://doi.org/10.1017/CBO9781139174411.

Lee, Y.H., Schmidt, P., 1993. A production frontier model with flexible temporal variation in technical efficiency. In Fried, H. O., Lovell, C.A.K., Schmidt, S.S., eds., The measurement of productive efficiency: Techniques and applications. 237-255. Oxford University Press, 1993.

Meeusen, W., van Den Broeck, J., 1977. Efficiency Estimation from Cobb-Douglas Production Functions with Composed Error. International Economic Review, Vol. 18, No. 2 (Jun., 1977), pp. 435-444. Published By: Wiley. http://dx.doi.org/10.2307/2525757.

Merkert, R., Hensher, D.A., 2011. The impact of strategic management and fleet planning on airline efficiency - a random effects Tobit model based on DEA efficiency scores. Transport. Res. Pol. Pract. 45 (7), 686-695.

Ngo, T., Tsui, K.W.H., 2022. Estimating the confidence intervals for DEA efficiency scores of Asia-Pacific airlines. Operational Research, Volume 22, pages 3411-3434, (2022). Published online: 7 August 2021. DOI:10.1007/s12351-021-00667-w.

Tsionas, M.G., Chen, Z., Wanke, P., 2017. A structural vector autoregressive model of technical efficiency and delays with an application to Chinese airlines. Transportation Research Part A: Policy and Practice. Elsevier, vol. 101(C), pages 1-10. DOI: 10.1016/j.tra.2017.05.003.

Xu, Y., Park, Y.S., Park, J.D., Cho, W., 2021. Evaluating the environmental efficiency of the U.S. airline industry using a directional distance function DEA approach. Journal of Management Analytics, Volume 8, 2021 - Issue 1. Published Online: 21 Oct 2020. DOI:10.1080/23270012.2020.1832925.